\documentclass[onecollarge,natbib]{svjour2} 
\bibpunct{[}{]}{;}{n}{}{,} 

\smartqed  
\usepackage{graphicx,amsmath,enumerate}

\usepackage{bm,hyphenat,xspace}
\usepackage{graphicx,epsfig}

\newcommand{\He}{${}^4\mathrm{He}$ }

\journalname{Few-body systems}
\begin{document}

\title {Momentum-space calculation of ${}^4${He} triatomic system
with realistic potential
}

\author{A. Deltuva} 
\institute{ A. Deltuva  \at
Institute of Theoretical Physics and Astronomy, 
Vilnius University, A. Go\v{s}tauto 12, LT-01108 Vilnius, Lithuania \\
              \email{arnoldas.deltuva@tfai.vu.lt} } 

\date{Received: date / Accepted: date}

 \maketitle

\begin{abstract}
The system of three ${}^4\mathrm{He}$ atoms with realistic interactions is studied
in the momentum-space framework. It is demonstrated that  
short and long-range  difficulties 
encountered in the coordinate-space calculations can be reliably resolved  in the momentum-space calculations.
Well-converged and accurate results are obtained for the ground and excited trimer energies, atom-dimer 
scattering length,  phase shifts, inelasticity parameters,
and elastic and breakup cross sections. A significant correction to previously published results is
found for the elastic cross section at very low energy.
\keywords{ Transition operators \and three-particle scattering} 
\PACS{  34.50.-s \and 34.10.+x}
\end{abstract}

\section{Introduction}

Systems of \He atoms have attracted significant interest, 
both experimentally and theoretically.
The main reason is a special feature of \He few-atom clusters, namely,
an extremely weak binding of few-body ground states.
This weakness of binding
is responsible for the manifestation of phenomena such as the Efimov
physics or superfluidity; see Ref.~\cite{kolganova:09a} for an overview.
The understanding of these phenomena requires precise theoretical calculations
of the properties of cold \He few-body systems;
up to now such calculations with realistic interaction models
predominantly have been performed in various coordinate-space frameworks.
Three-atom bound (trimer) and scattering states have been calculated 
solving Faddeev equations 
\cite{kolganova:09a,roudnev:03a,lazauskas:he}
or using adiabatic hyperspherical approach \cite{suno:08a},
while tetramer  properties have been obtained
solving the Faddeev-Yakubovsky equations \cite{lazauskas:he},
using the variational Gaussian expansion method \cite{hiyama:12a}
or the hyperspherical Monte-Carlo approach  \cite{blume:00a}.
The latter was extended to more than four atoms and provided
estimates for scattering lengths, 
whereas Ref.~\cite{lazauskas:he}
produced results also for the atom-trimer scattering at finite energy.

The dynamics of  cold \He atoms is essentially governed by the local
two-atom potential; there exist a number of realistic
parametrizations \cite{aziz}.
Main features of those realistic potentials are long attractive
van der Waals tail and a very strong repulsion at short distances.
Especially the latter causes difficulties in calculations of systems
containing three or more  \He atoms.
Indeed, even a special form of the differential
three-particle Faddeev equations 
with hard-core potentials has been developed and applied for
the study of the three-  \cite{kolganova:09a}
and four-atom systems \cite{lazauskas:he}.
Another difficulty, encountered in the scattering calculations,
is a slow convergence of the results with the employed
grid size \cite{kolganova:09a,roudnev:03a,lazauskas:he}.
To avoid these difficulties, simplified dynamic models
based of short-range soft or separable potentials have been invented
and used in coordinate- \cite{kievsky:11a} 
and momentum-space calculations \cite{shepard:07a}; despite their
simplicity such models are quite reliable for the study of Efimov physics.

Given the success of momentum-space three-nucleon scattering calculations
\cite{gloeckle:96a,deltuva:03a},
one may raise the question whether the above difficulties 
in the description of atomic \He systems with realistic potentials can
be resolved in the framework of  momentum-space integral equations
for the transition operators. Therefore the goal of the present
work is to study the \He trimer and  three-atom scattering processes
employing the momentum-space methods. Since the \He triatomic  system has 
already been investigated using several coordinate-space methods
\cite{kolganova:09a,roudnev:03a,lazauskas:he,suno:08a,hiyama:12a},
in this work I will concentrate on demonstrating the reliability
of  momentum-space calculations with realistic potentials
and providing highly accurate
benchmark results.  The considered observables are the trimer binding energy,
 atom-dimer scattering length, phase shifts, inelasticity parameters, 
inclusive cross sections 
as well as differential cross section for the atom-dimer breakup;
the latter is most difficult to obtain in the
coordinate-space framework and, to the best of my knowledge,
has not yet been calculated.

Among the variety of realistic atomic \He potentials, the one by 
Aziz and Slaman, called LM2M2 \cite{aziz}, 
is probably most widely used and therefore is an optimal
choice for a benchmark calculation; 
having van der Waals tail and  very strong short-range repulsion
its behaviour is characteristic for all realistic \He potentials.
The fact that LM2M2 does not include
the retardation correction which may be quite significant
\cite{suno:08a} is irrelevant for demonstrating the abilities of the
method. As in most previous calculations the value of  \He mass $m$
given by the relation $\hbar^2/m = 12.12$ K\AA$^2$ is used.

In Sec.~\ref{sec:4bse} I describe the employed three-atom 
equations and the technical framework. 
In Sec.~\ref{sec:res} I present results for the \He trimer 
and atom-dimer scattering.
I summarize in  Sec.~\ref{sec:sum}.

\section{Integral equations for three-atom bound state and 
scattering  \label{sec:4bse}}

In order to apply the momentum-space techniques to the 
three-body problem one has to formulate it in the integral representation.
In this way the asymptotic boundary conditions are implicitly
built into the form of equations but lead to kernels with integrable singularities.
Considering the three \He atoms as identical bosons  it is most 
convenient to use the symmetrized form of Faddeev equations
in the Alt-Grassberger-Sandhas (AGS) version
\cite{alt:67a};
formally they coincide with the ones employed for the 
fermionic three-nucleon system \cite{gloeckle:96a,deltuva:03a}.
The only difference is in the properties of the basis states,
namely, they must be symmetric under the exchange of particles
within pair (12) which is chosen as a representative pair.
The potential $v$ acting within this pair 
 in the integral formulation is summed up into the 
respective two-particle transition matrix using the
Lippmann-Schwinger equation
\begin{equation} \label{eq:t}
t = v + v G_0 t,
\end{equation}
with  $G_0 = (E+i0-H_0)^{-1}$ being the free Green's function of the
three-particle system with energy $E$ and kinetic energy operator $H_0$.
 The required full symmetry of the 
three-boson system is ensured by the  operator 
$P =  P_{12}\, P_{23} + P_{13}\, P_{23}$  where $P_{ab}$ is the
permutation operator of particles $a$ and $b$.
With these definitions the Faddeev amplitude $|\psi\rangle$
 of the trimer wave function $|B\rangle = (1+P)|\psi\rangle$
is the solution of the homogeneous Faddeev equation
\begin{equation} \label{eq:fa}
|\psi\rangle =  G_0 t P |\psi\rangle
\end{equation}
at $E = -E_t$ with $E_t$..being the binding energy.

In the case of the atom-dimer scattering the AGS equations are formulated 
for three-particle transition operator
\begin{equation} \label{eq:U3}
U =  P G_0^{-1} + P  t G_0  U;
\end{equation}
it contains the full information about the scattering process.
The amplitude for the elastic scattering is obtained from the 
on-shell matrix element $\langle \phi \vec{q}_f|U|\phi \vec{q}_i \rangle$
where $|\phi \vec{q}_j \rangle$ is the atom-dimer channel state given
by the product of the dimer wave function $|\phi \rangle$
and plane wave with the relative atom-dimer momentum $\vec{q}_j$.
The latter is related to the available energy $E$ as
$E = -e_d + q_j^2/2\mu$ with $e_d$ being the dimer binding energy and 
$\mu = 2m/3$ the reduced mass.
The breakup amplitudes $\langle \vec{p} \vec{q}|U_0|\phi \vec{q}_i \rangle$,
with $ \vec{p}$ and $ \vec{q}$ being the Jacobi momenta  of three free particles
in the final state, are obtained as on-shell matrix elements of  the breakup operator
\begin{equation} \label{eq:U0}
U_0 =  (1+P) G_0^{-1} + (1+P)  t G_0  U.
\end{equation}
The above equations are solved in the momentum-space partial-wave basis
$|pq(Ll)JM\rangle$ where $p$ and $q$ are magnitudes of the Jacobi momenta
and $L$ ($l$) is the orbital angular momentum for the relative motion
within the pair (between the pair and spectator), coupled to the total angular
momentum $J$ with the projection $M$. The solution technique employs the
same numerical methods as in the three-nucleon scattering calculations, i.e.,
real-axis integration with subtraction and special weights, spline interpolation,
and  Pad\'{e} summation.
Further details can be found in Refs.~\cite{deltuva:03a,chmielewski:03a}.

However, prior to solving Eqs.~(1-3) one has to transform the coordinate space  potential
$V(r)$ to the momentum-space partial wave representation
\begin{equation} \label{eq:v}
\langle p'L | V | p L \rangle = \frac{2}{\pi} \int_0^\infty
j_L(p'r)V(r)j_L(pr) \, r^2 dr,
\end{equation}
where $j_L(x)$ is the spherical Bessel function of the order $L$.
Given the two difficulties encountered in the coordinate-space calculations, 
the essential questions for the momentum-space calculations are:
(i) To what extent the hard core region is important and causes difficulties;
(ii) At what distances the van der Waals tail becomes irrelevant. 
In the latter case, given the fact that the  integral form of scattering equations implicitly
incorporates the asymptotic boundary conditions, one may expect significant reduction
in the distance $r_{\rm max}$ up to which the $r$-space integration needs to be performed.
In order to answer these questions I introduce a modified  potential
\begin{gather} \label{eq:vr}
\langle p'L | v | p L \rangle = \frac{2}{\pi} \int_0^{r_{\rm max}} j_L(p'r)  \,  
\{  \Theta(r_{\rm min}-r) V(r_{\rm min})[2 - e^{\kappa(r_{\rm min}-r)} ] 
  +   \Theta(r-r_{\rm min})V(r) \}
  j_L(pr) \, r^2 dr,
\end{gather}
where the step function $\Theta(x)$  equals  1 if $x > 0$ and 0 otherwise.
The parameter $\kappa$ is chosen as $\kappa = V'(r_{\rm min})/V(r_{\rm min})$
thereby ensuring the continuity of the derivative  $V'(r)$ and the 
smoothness of the modified potential at $r=r_{\rm min}$. 
Note that in the limit $r_{\rm min} \to 0$, $r_{\rm max} \to \infty$ one has
$\langle p'L | v | p L \rangle = \langle p'L | V | p L \rangle $.
The $r$-integration in Eq.~(\ref{eq:vr}) is performed numerically using 
the standard Gaussian quadrature. As the integral is one-dimensional, one may
easily include several thousands or even more grid points and obtain very accurate
results for $\langle p'L | v | p L \rangle$. These matrix elements via Eqs.~(1-4)
are  used to calculate various three-atom observables 
whose dependence on the parameters  $r_{\rm min}$ and $r_{\rm max}$ is then investigated.
The findings are presented in the next section.

\section{Results \label{sec:res}}
 
I start by demonstrating the convergence
of obtained results with $r_{\rm min}$ and $r_{\rm max}$. 
The example observables are the dimer binding energy $e_d$,
the trimer ground and excited state binding energies $E_t$ and $E_{t*}$, the atom-dimer scattering
length $a_0$, and phase shifts $\delta_l$ and inelasticity parameters $\eta_l$ of 
the atom-dimer scattering
at the kinetic center-of-mass (c.m.) energy $E_k = 40$ mK. The relation between these scattering parameters and
partial-wave on-shell elements $\langle \phi {q}_i l|U|\phi {q}_i l\rangle$
of the transition operator is given by
\begin{subequations} \label{eq:res}
\begin{align}
a_0 = {} & \pi \mu \langle \phi 0 0|U|\phi 0 0\rangle, \\
\eta_l e^{2i\delta_l} = {} & 1- 2\pi i \mu q_i \langle \phi q_i l|U|\phi q_i l\rangle.
\end{align}
\end{subequations}
The results are converged with respect to the number of grid points for the Jacobi momenta
$p$ and $q$, taking about 80 to 100 points for each momentum, and with respect to the number of partial
waves, including states with $L \le 12$. In the latter case the observed convergence pattern
for  $E_t$, $E_{t*}$, and $a_0$
is in full agreement with the coordinate-space Faddeev results of Ref.~\cite{lazauskas:he}.
To achieve the   given accuracy for $\delta_l$ and  $\eta_l$ it is sufficient to include
the states with $L \le 10$.

\begin{table}[h]
\begin{tabular}{*{2}{r}*{8}{c}} $r_{\rm min}$ & $r_{\rm max}$ & $e_{d}$ & 
  $E_t$ & $E_{t*}$ & $a_0$ & $\delta_0$ & $\eta_0$ & $\delta_1$ & $\eta_1$
\\  \hline
1.5 &  25 & 1.3010 & 126.40 & 2.267 & 115.32 &  24.92  &  0.8281  & -73.78  &  0.7671 \\
1.5 &  50 & 1.3034 & 126.40 & 2.271 & 115.22 &  24.93  &  0.8283  & -73.78  &  0.7675 \\
1.5 &  75 & 1.3035 & 126.40 & 2.271 & 115.21 &  24.93  &  0.8283  & -73.78  &  0.7676 \\
1.5 & 100 & 1.3035 & 126.40 & 2.271 & 115.21 &  24.93  &  0.8283  & -73.78  &  0.7676 \\
1.5 & 200 & 1.3035 & 126.40 & 2.271 & 115.21 &  24.93  &  0.8283  & -73.78  &  0.7676 \\
\hline
0.0  & 100 & 1.3035  &  126.40 &  2.271  &  115.21 &   24.93  & 0.8283 &  -73.78 &  0.7676  \\
0.5  & 100 & 1.3035  &  126.40 &  2.271  &  115.21 &   24.93  & 0.8283 &  -73.78 &  0.7676  \\
1.0  & 100 & 1.3035  &  126.40 &  2.271  &  115.21 &   24.93  & 0.8283 &  -73.78 &  0.7676  \\
1.5  & 100 & 1.3035  &  126.40 &  2.271  &  115.21 &   24.93  & 0.8283 &  -73.78 &  0.7676  \\
1.7  & 100 & 1.3035  &  126.40 &  2.271  &  115.21 &   24.93  & 0.8283 &  -73.78 &  0.7676  \\
1.8  & 100 & 1.3041  &  126.41 &  2.272  &  115.22 &   24.93  & 0.8283 &  -73.78 &  0.7677  \\
1.9  & 100 & 1.3093  &  126.51 &  2.278  &  115.36 &   24.93  & 0.8284 &  -73.76 &  0.7687  \\
2.0  & 100 & 1.3434  &  127.13 &  2.321  &  116.23 &   24.95  & 0.8290 &  -73.64 &  0.7754  \\
\hline
\end{tabular}
\caption{ \label{tab:res-r}
Observables in three-atom system as functions of the coordinate-space cut-off parameters  $r_{\rm min}$ and $r_{\rm max}$
 that are given in units of \AA. 
Dimer binding energy $e_d$, trimer binding energy for ground and excited state $E_t$ and $E_{t*}$
(all in units of mK), atom-dimer scattering
length $a_0$ (in units of \AA), and phase shifts $\delta_l$ (in degrees) and inelasticity parameters $\eta_l$ for 
the atom-dimer scattering at $E_k = 40$ mK are listed.
}
\end{table}

By inspecting the Table \ref{tab:res-r} one may conclude that, within the given accuracy, 
the results become independent of $r_{\rm min}$ for $r_{\rm min} \le 1.7$ \AA{}
and independent of $r_{\rm max}$ for $r_{\rm max} \ge 75$ \AA. For comparison, 
to get $a_0$ with a comparable precision when solving coordinate-space equations,
the integration up to distances well above 1000 \AA{} was needed  
\cite{kolganova:09a,roudnev:03a,lazauskas:he}.
Thus, the integral formulation of the scattering problem with implicitly imposed boundary conditions 
allows indeed for a significant reduction in the maximal interparticle distance.
The converged values agree well with the most accurate ones 
provided in the literature \cite{lazauskas:he,hiyama:12a,roudnev:11a} for 
$E_t$, $E_{t*}$, and  $a_0$; this comparison is presented in Table~\ref{tab:cmp}.
The expectation values for internal kinetic energies of trimer states
$\langle K_t \rangle$ and $\langle K_{t*} \rangle$ show good agreement as well.
The hyperspherical Monte-Carlo approach  \cite{blume:00a} is less accurate for $a_0$,
while the differences between the results of \cite{kolganova:09a} and 
\cite{lazauskas:he,hiyama:12a,roudnev:11a} are due to the limitation $L \le 4$ used 
in Ref.~\cite{kolganova:09a}.

\begin{table}[h]
\begin{tabular}{{l|}*{5}{l}}  Reference & 
  $E_t$ & $\langle K_t \rangle$ & $E_{t*}$ & $\langle K_{t*} \rangle$ & $a_0$ 
\\  \hline
This work          & 126.40 & 1660.5 & 2.271 & 122.13 & 115.21   \\
\cite{lazauskas:he}& 126.39 & 1658   & 2.268 & 122.1  & 115.2  \\
 \cite{hiyama:12a} & 126.40 & 1660.4 & 2.2706 & 122.15 & \\
\cite{roudnev:03a,roudnev:05a} & 126.41 &  & 2.271 &  & 115.4 \\
\cite{roudnev:11a} & 126.394 &  & 2.2711 &  & 115.22 \\
\cite{kolganova:09a} & 125.9 & & 2.28 & & 117.0 \\
\cite{blume:00a} & 125.5 & & 2.19 & & 126 \\
\hline
\end{tabular}
\caption{ \label{tab:cmp}
Trimer properties and atom-dimer scattering length as obtained in different works.
$a_0$ is  given in units of \AA{} while other quantities in units of mK.
}
\end{table}

Regarding the atom-dimer scattering at finite energies, the  available
results are scarce and lack consistency. For example, the $S$-wave ($l=0$) phase shifts  $\delta_0$ 
  show some disagreements
between Refs.~\cite{kolganova:09a,roudnev:03a,suno:08a}.
 $\delta_0$ values below breakup threshold are tabulated in Ref.~\cite{roudnev:03a},
however, it appears that Ref.~\cite{roudnev:03a} uses 
 a nonstandard convention  which makes the comparison ambiguous. Indeed, while the 
standard relation between $a_0$ and $\delta_0$ reads
$\lim_{q_i \to 0} \tan{\delta_0}/q_i = -a_0$, an inspection of $\delta_0$ results from 
Ref.~\cite{roudnev:03a} reveals the relation
$\lim_{q_i \to 0} \sqrt{4/3}\tan{\delta_0}/q_i = -a_0$.
This may also explain why $\delta_0$ values from   Ref.~\cite{roudnev:03a}
are higher than those of Ref.~\cite{kolganova:09a}.
To sort these discrepancies I present in Table \ref{tab:ph}
highly accurate momentum-space results
for atom-dimer phase shifts $\delta_l$ and inelasticity parameters $\eta_l$ up to $l=5$.
A broad range of kinetic energies in the c.m. system
is considered, ranging from 0.1 mK to 40 mK which is well above the breakup
threshold of 1.3035 mK.

\begin{table}[h]
\begin{tabular}{*{13}{r}} 
$E_k$  & $\delta_0$ & $\eta_0$ & $\delta_1$ & $\eta_1$ & $\delta_2$ & $\eta_2$ & $\delta_3$ & $\eta_3$ & $\delta_4$ & $\eta_4$ & $\delta_5$ & $\eta_5$ \\
       \hline
   0.1  &  158.05 &  1.0000 &  -2.39 &  1.0000 &   0.11 &  1.0000 &  -0.00 &  1.0000 &   0.00 &  1.0000 &  -0.00 &  1.0000  \\
   0.2  &  149.11 &  1.0000 &  -5.40 &  1.0000 &   0.44 &  1.0000 &  -0.03 &  1.0000 &   0.00 &  1.0000 &  -0.00 &  1.0000  \\
   0.4  &  137.11 &  1.0000 & -10.76 &  1.0000 &   1.51 &  1.0000 &  -0.17 &  1.0000 &   0.02 &  1.0000 &  -0.00 &  1.0000  \\
   1.0  &  116.71 &  1.0000 & -21.81 &  1.0000 &   4.91 &  1.0000 &  -0.96 &  1.0000 &   0.20 &  1.0000 &  -0.04 &  1.0000  \\
   2.0  &   99.04 &  0.9994 & -32.32 &  1.0000 &   8.12 &  0.9995 &  -2.23 &  1.0000 &   0.62 &  1.0000 &  -0.18 &  1.0000  \\
   4.0  &   80.79 &  0.9989 & -43.48 &  0.9996 &  10.33 &  0.9833 &  -3.60 &  0.9981 &   1.17 &  0.9997 &  -0.42 &  1.0000  \\
  10.0  &   57.52 &  0.9853 & -57.30 &  0.9840 &  11.27 &  0.8878 &  -4.50 &  0.9775 &   1.54 &  0.9943 &  -0.67 &  0.9987  \\
  20.0  &   40.90 &  0.9319 & -66.05 &  0.9187 &  11.41 &  0.7696 &  -4.30 &  0.9364 &   1.43 &  0.9795 &  -0.70 &  0.9935  \\
  40.0  &   24.93 &  0.8283 & -73.78 &  0.7676 &  11.84 &  0.6610 &  -3.43 &  0.8782 &   1.26 &  0.9527 &  -0.59 &  0.9807  \\
\hline
\end{tabular}
\caption{ \label{tab:ph}
Atom-dimer phase shifts   $\delta_l$ (in degrees) and inelasticity parameters $\eta_l$ as functions  of the   kinetic energy $E_k$ (in mK) in the c.m. system.
}
\end{table}

\begin{figure}[!]
\includegraphics[scale=0.75]{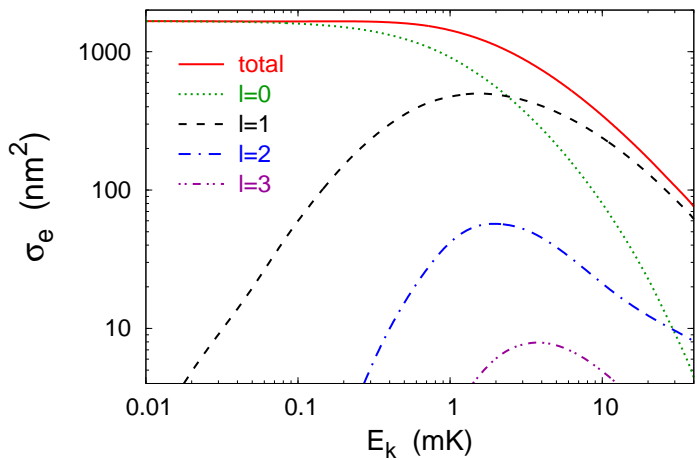} 
\caption{\label{fig:se} (Color online)
Atom-dimer total elastic cross section and its partial-wave
contributions as functions the c.m. kinetic energy.
Contributions with $l \ge 4$ remain well below 2 nm$^2$.
}
\end{figure}

In addition, elastic cross section and breakup rate results, albeit with a different potential, are given in 
 Ref.~\cite{suno:08a}. However, the elastic cross section  $\sigma_e$ in Ref.~\cite{suno:08a}
 shows a very strange energy dependence in
 partial waves with $l>0$: at $E_k = 0.001$ mK the partial cross sections $\sigma_e(l)$ are large
 but decrease with increasing energy 
 in contrast to the standard near-threshold behaviour, and, furthermore,  $\sigma_e(l)$ show the trend to increase with increasing $l$
 such that 
one can even question the partial-wave convergence. Such a strange behaviour is absent in the momentum-space results
 displayed in Fig.~\ref{fig:se}. Below 0.1 mK the elastic cross section is almost constant and strongly dominated by the $l=0$ state,
 whereas very small $l>0$ contributions increase with increasing energy and decrease with increasing $l$. This is not unexpected given the energy
 evolution of phase shifts $\delta_l$ in Table \ref{tab:ph} that show a similar trend.
 One might suspect that larger distances in Eq.~(\ref{eq:vr})
become more important at very low energies. To verify
  the convergence of the results I performed additional calculations with $r_{\rm max}$ up to 4000 \AA,
  but found no changes to the results of  Fig.~\ref{fig:se}.
 Note, however, that due to the van der Waals tail $1/r^6$ the phase shifts $\delta_l$ and partial cross sections $\sigma_e(l)$
 follow the standard near-threshold behaviour $\delta_l \sim q_i^{2l+1}$ and $\sigma_e(l) \sim E_k^{2l}$ only
 for low $l$. 
  
\begin{figure}[!]
\includegraphics[scale=0.75]{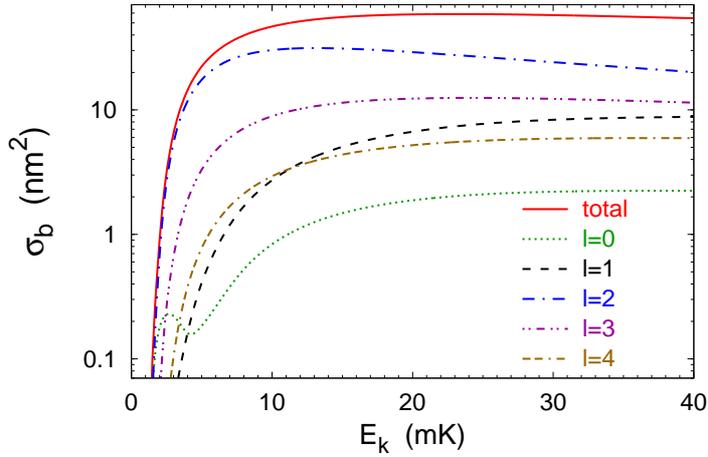} 
\caption{\label{fig:sb} (Color online)
Atom-dimer total breakup cross section and its most sizable partial-wave
contributions as functions the c.m. kinetic energy.}
\end{figure}

Regarding the total breakup cross section $\sigma_b$, there is a good qualitative agreement between the present 
momentum-space results displayed in Fig.~\ref{fig:sb} and those of Ref.~\cite{suno:08a} extracted from the breakup rate $\sigma_b q_i/\mu$. 
Above $E_k = 3$ mK in both cases   $\sigma_b$ is dominated by initial states $l=2$ and 3 whereas $l=0$ contribution is lower by 
an order of magnitude. Furthermore, the latter exhibits a local minimum around $E_k = 4$ mK seen in Fig.~\ref{fig:sb} 
as well as in Ref.~\cite{suno:08a}.

\begin{figure}[!]
\includegraphics[scale=0.75]{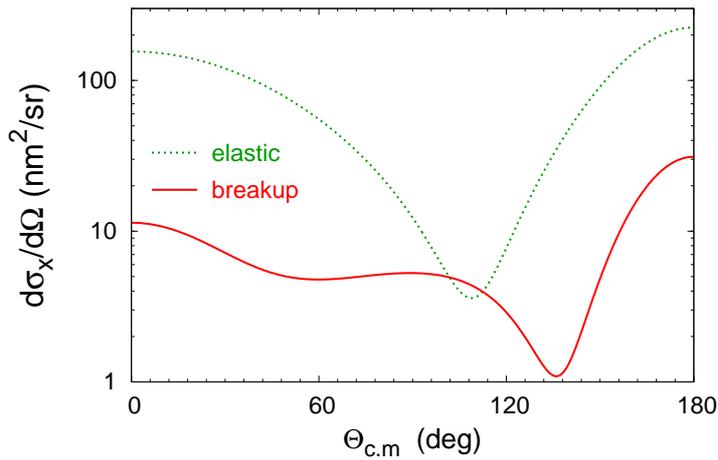} 
\caption{\label{fig:ds} (Color online)
Differential cross sections for the 
atom-dimer elastic scattering (dotted curve) and breakup (solid curve)
at $E_k = 5$ mK 
as functions of the atom scattering angle in the c.m. frame.}
\end{figure}

\begin{figure}[!]
\includegraphics[scale=0.66]{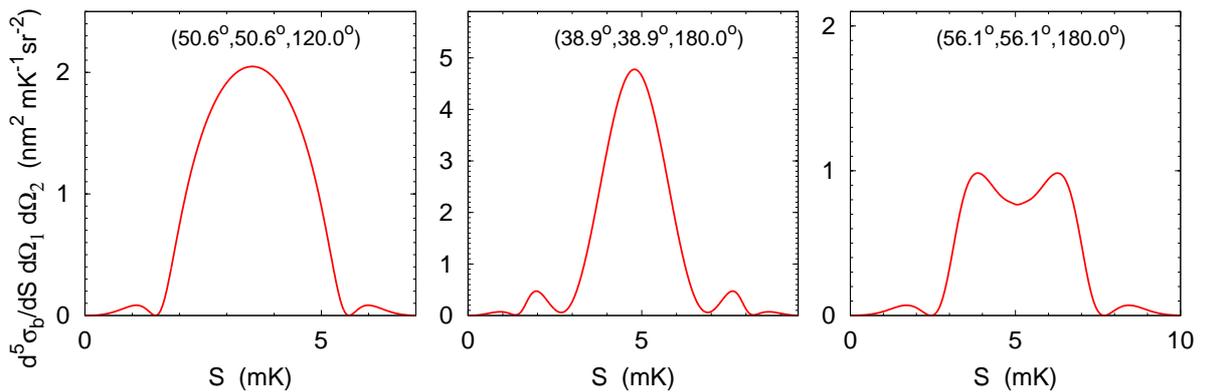} 
\caption{\label{fig:d5s} (Color online)
Differential cross sections for the 
atom-dimer  breakup at $E_k = 5$ mK 
as functions of the energy parameter $S$ in the space star (left), quasi free scattering (middle),
and collinear (right) kinematical configurations.}
\end{figure}

The rate for the time reversed reaction, the three-atom recombination, is simply proportional to  
$\sigma_b$ with a kinematic factor \cite{deltuva:13a} and therefore is not shown separately. 
These observables, owing to their inclusive character, can be obtained even without explicit calculation 
of breakup or recombination operator $U_0$ but just using the optical theorem for $U$. 
Exclusive or semi-inclusive observables are more difficult to calculate, and the  transition
operator momentum-space framework may be the most appropriate method for such calculations,
much like in the three-nucleon physics \cite{gloeckle:96a,chmielewski:03a,deltuva:05d}.
To demonstrate its abilities I present in Fig.~\ref{fig:ds} the angular distribution of the 
semi-inclusive differential cross section $d\sigma_b/d\Omega$ and in Fig.~\ref{fig:d5s}
the fully exclusive fivefold differential cross section $d^5\sigma_b/d\Omega_1 d\Omega_2 dS$,
where $\Omega_j$ is the solid angle of the $j$-th atom, given by polar and azimuthal angles $(\Theta_j,\varphi_j)$,
and $S$ is the distance along the kinematical curve, routinely employed in three-nucleon physics \cite{gloeckle:96a,chmielewski:03a}.
The energy $E_k = 5$ mK is chosen such that the ratio $E_k/e_d$ is nearly the same as for the nucleon lab energy
of 13 MeV in the nucleon-deuteron breakup where numerous calculations exist \cite{gloeckle:96a,chmielewski:03a,deltuva:05d}.

Since higher three-body partial waves contribute to breakup and the final phase space is of higher dimension,
the breakup differential cross section $d\sigma_b/d\Omega$  exhibits a more complicated angular dependence in the breakup
as compared to the elastic one $d\sigma_e/d\Omega$  that is also shown in Fig.~\ref{fig:ds}

The fully exclusive fivefold differential cross section $d^5\sigma_b/d\Omega_1 d\Omega_2 dS$ is shown in 
Fig.~\ref{fig:d5s} for three special kinematical configurations 
$(\Theta_1, \Theta_2, \varphi_2-\varphi_1) = (50.6^\circ,50.6^\circ,120.0^\circ)$, 
$(38.9^\circ,38.9^\circ,180.0^\circ)$, and $(56.1^\circ,56.1^\circ,180.0^\circ)$, that 
are called the space star, quasi free scattering (QFS), and collinear configuration, respectively 
\cite{gloeckle:96a,deltuva:05d}. It is quite interesting to compare these results with the ones
for the nucleon-deuteron breakup at 13 MeV in the corresponding configurations
as given in Ref.~\cite{deltuva:05d}.
Although there are some similarities,
e.g., the existence of the QFS peak and of the local minimum at the collinear point (around $S= 5$ mK in both cases),
there are also significant differences, namely, (i) the reduction (for 
QFS and collinear) or absence (for space star) of other peaks with large cross section, and (ii)
the presence of local minima with very small
$d^5\sigma_b/d\Omega_1 d\Omega_2 dS$ for $S<2.5$ mK and $S_{\rm max}-S<2.5$ mK.  Possible  explanations for these differences
are following:
In the case (i) the two-nucleon ${}^1S_0$ virtual state is responsible, at least partially, for the  differential 
cross section peaks observed in the nucleon-deuteron breakup; there is no similar state in the \He atomic system.
In the case (ii) the $d^5\sigma_b/d\Omega_1 d\Omega_2 dS$ may get modulated by
the nodes in the momentum-space representation of the \He dimer wave function
that are more numerous than for the deuteron.

\section{Summary \label{sec:sum}}

I performed bound state and scattering calculations in the system of three \He atoms using realistic
two-atom potential. Exact integral equations for the Faddeev amplitude and transition operators were
solved in the momentum-space partial-wave framework. I demonstrated that two technical
complications present in the coordinate-space calculations, i.e., the hard core and the need for 
a very extended grid, create no major difficulties in the momentum-space calculations,
and well-converged results are obtained.
Predictions for the ground and excited trimer energies and the atom-dimer 
scattering length agree well with the most accurate coordinate-space results obtained by other authors.
I also presented high-precision benchmark results for the atom-dimer phase shifts, inelasticity parameters,
and elastic and breakup cross sections over a broad energy range.
An important finding in the case of low-energy elastic scattering is that the unusual energy and angular momentum
dependence of partial cross sections $\sigma_e(l)$ predicted in Ref.~\cite{suno:08a} is not confirmed;
instead, a strong dominance of the $l=0$ wave  is found in the present work
while $l\ge 1$ contributions become practically negligible for energies below 0.1 mK.
Despite this disagreement, the results for the total breakup cross section in both calculations are
consistent.
The semi-inclusive and fully  exclusive breakup  differential cross sections are calculated for the
first time and compared with those  in the nucleon-deuteron breakup.
It is conjectured that \He dimer wave function nodes may be related to the minima of the
fivefold differential cross section.


\end{document}